\documentclass[5p,twocolumn,sort&compress]{elsarticle}
\usepackage{amssymb,amsmath}
\usepackage{url}
\usepackage{etoolbox}
\usepackage{graphicx}
\usepackage{tikz}
\usetikzlibrary{patterns}
\usepackage{siunitx}
\usepackage[utf8]{inputenc}
\usepackage{dblfloatfix}
\usepackage{chngcntr}

\journal{Computer Physics Communications}

\begin{document}
%\sisetup{mode=text,range-phrase = {\text{~to~}}}
\sisetup{mode=text,range-phrase = -}

\begin{frontmatter}
\title{Massively parallel multicanonical simulations}

\author[leipzig]{Jonathan Gross}
\ead{gross@itp.uni-leipzig.de}
\author[leipzig,lllc]{Johannes Zierenberg\fnref{fn1}}
\ead{zierenberg@itp.uni-leipzig.de}
\author[lllc,coventry]{Martin Weigel}
\ead{Martin.Weigel@coventry.ac.uk}
\author[leipzig,lllc]{Wolfhard Janke}
\ead{janke@itp.uni-leipzig.de}

\address[leipzig]{Institut f\"ur Theoretische Physik, Universit\"at Leipzig, Postfach 100920, D--04009 Leipzig, Germany}
\address[lllc]{Doctoral College for the Statistical Physics of Complex Systems, Leipzig-Lorraine-Lviv-Coventry $({\mathbb L}^4)$, D-04009 Leipzig, Germany}
\address[coventry]{Applied Mathematics Research Centre, Coventry University, Coventry CV1 5FB, England}

\fntext[fn1]{Present address: Max Planck Institute for Dynamics and Self-Organization, Am Fassberg 17, D-37077 Göttingen, Germany}
\begin{abstract}
  
  Generalized-ensemble Monte Carlo simulations such as the multicanonical method and
  similar techniques are among the most efficient approaches for simulations of
  systems undergoing discontinuous phase transitions or with rugged free-energy
  landscapes. As Markov chain methods, they are inherently serial computationally. It
  was demonstrated recently, however, that a combination of independent simulations
  that communicate weight updates at variable intervals allows for the efficient
  utilization of parallel computational resources for multicanonical
  simulations. Implementing this approach for the many-thread architecture provided
  by current generations of graphics processing units (GPUs), we show how it can be
  efficiently employed with of the order of $10^4$ parallel walkers and beyond, thus
  constituting a versatile tool for Monte Carlo simulations in the era of massively
  parallel computing. We provide the fully documented source code for the approach
  applied to the paradigmatic example of the two-dimensional Ising model as starting
  point and reference for practitioners in the field.

\end{abstract}


\end{frontmatter}

{\bf PROGRAM SUMMARY}

\begin{small}
\noindent
{\em Manuscript Title:} Massively parallel multicanonical simulations\\
{\em Authors:} Jonathan Gross, Johannes Zierenberg, Martin Weigel, Wolfhard Janke \\
{\em Program Title:} cudamuca \\
{\em Journal Reference:}                                      \\
  %Leave blank, supplied by Elsevier.
{\em Catalogue identifier:}                                   \\
  %Leave blank, supplied by Elsevier.
{\em Licensing provisions:}   Creative Commons Attribution license (CC BY 4.0)                                 \\
  %enter "none" if CPC non-profit use license is sufficient.
{\em Programming language:} C, CUDA                                \\
{\em Computer:} System with an NVIDIA CUDA enabled GPU \\
  %Computer(s) for which program has been designed.
{\em Operating system:}  Linux, Windows, macOS   \\
  %Operating system(s) for which program has been designed.
{\em RAM:} $O(\SI{100}{MB})$ depending on system size (about \SI{200}{MB} for L=32)\\
  %RAM in bytes required to execute program with typical data.
{\em Number of processors used:} 1 GPU \\
  %If more than one processor.
% {\em Supplementary material:}                                 \\
  % Fill in if necessary, otherwise leave out.
  % Please give some freely chosen keywords that we can use in a
  % cumulative keyword index.
{\em Classification:} 23   \\
  %Classify using CPC Program Library Subject Index, see (
  % http://cpc.cs.qub.ac.uk/subjectIndex/SUBJECT_index.html)
  %e.g. 4.4 Feynman diagrams, 5 Computer Algebra.
{\em External routines/libraries:} NVIDIA CUDA Toolkit 6.5 or newer     \\
  % Fill in if necessary, otherwise leave out.
% {\em Subprograms used:}                                       \\
  %Fill in if necessary, otherwise leave out.
{\em Nature of problem: }%
The program determines weights for a multicanonical simulation of the 2D Ising model
to result in a flat energy histogram.
A final production run with these weights provides an estimate of the density of
states of the model.
% Describe the nature of the problem here.
\\
{\em Solution method: }%
The code uses a parallel variant of the multicanonical method employing many parallel
walkers that accumulate a common histogram. The resulting histogram is used
to determine the weight function for the next iteration. Once the iteration has
converged, simulations visit all possible energies with the same probability.
% Describe the method solution here.
\\
{\em Restrictions: }%
The system size and size of the population of replicas
are limited depending on the memory of the GPU device used. \\
% Describe any restrictions on the complexity of the problem here.
% {\em Unusual features:}\\
  %Describe any unusual features of the program/problem here.
% {\em Additional comments:}\\
Code repository at \url{https://github.com/CQT-Leipzig/cudamuca}.
\\
  %Provide any additional comments here.
{\em Running time:} Depends on system size (approx. \SI{40}{s} for L=32 on a
Tesla K20m)
\\

\end{small}
\setcounter{footnote}{0} % footnotes in text should start with 1

%%%%%%%%%%%%%%%%%%%%%%%%%%%%%%%%%%%%%%%%%%%%%%%%%%%%%%%%%%%%%%%%%%%%%%%%%%%%%%%%
%%%%%%%%%%%%%%%%%%%%%%%%%%%%%%%%%%%%%%%%%%%%%%%%%%%%%%%%%%%%%%%%%%%%%%%%%%%%%%%%
%%%%%%%%%%%%%%%%%%%%%%%%%%%%%%%%%%%%%%%%%%%%%%%%%%%%%%%%%%%%%%%%%%%%%%%%%%%%%%%%
\section{Introduction\label{secIntroduction}}

The age of regular rapid increases in serial computational performance of commodity
hardware came to an end about 15 years ago. Nonetheless, Moore's law, predicting an
exponential growth in the number of transistors in typical integrated circuits,
continues to hold~\cite{moore:65}. But now the additional transistors are used to
form further parallel computational units, or cores, instead of speeding up single
threads. Consequently, the number of cores available to researchers working on
cluster machines or supercomputers is growing rapidly, calling for the
parallelization of established computational approaches and algorithms or, where this
is not efficiently possible, the proposal of alternative solutions to the given
computational challenges~\cite{mccool:12}. In many cases, for instance for Molecular
Dynamics or Navier-Stokes simulations, such strategies will utilize domain
decompositions, where each processor or thread is assigned to the environment of a
subset of particles or cells. For Monte Carlo simulations the same approach works
well for simple local-update schemes, where the acceptance probability of a local
change is independent of any possible updates in distant areas. For non-local updates
such as cluster algorithms domain decompositions become more involved
\cite{weigel:10b}. Replica-exchange Monte Carlo and similar schemes, on the other
hand, are inherently parallel~\cite{hukushima:96a}. A further important class of
methods, namely generalized-ensemble simulations such as the
multicanonical~\cite{bergMuca, jankeMuca} and Wang-Landau~\cite{wang2001} approaches,
cannot be easily treated in the same way, however. There, the acceptance probability
for each move depends on the current value of a global variable, typically the
energy, thus effectively serializing all updates within the same Markov chain. The
natural alternative are approaches with several walkers working in parallel, be it
through independent runs for non-overlapping parameter (energy) ranges
\cite{schulz:03,weigel:11} or through a series of parallel full simulations with
rather infrequent synchronization of weights~\cite{vogel:13,zierenberg2013}.

Graphics processing units (GPUs) are by now rather well established as efficient,
massively parallel computational devices in scientific computing
\cite{owens:08}. They have been designed for the relatively low-level, highly
parallel task of rendering 3D graphics, i.e., with the goal of achieving high
performance under loads consisting of thousands of parallel rendering threads busy
with predictable arithmetic calculations. In contrast, modern CPU cores are designed
to provide good performance with an unpredictable, interactive serial or moderately
parallel load. Consequently, CPU dies feature a majority of transistors dedicated to
control logic and cache memories, whereas GPU dies are packed with arithmetic units,
resulting in significantly higher theoretical peak performances. While for good
results GPU codes need to take into account some fundamental design features of such
devices, it has been demonstrated that with relatively moderate coding effort GPU
based simulations can yield significant speed-ups relative to CPU based codes
compared at the same hardware cost~\cite{weigel2012}. In addition, GPU computing is
relatively energy efficient, a significant advantage in particular for supercomputing
and the quest for exascale computers. Thus, excellent performance of GPU codes could
be demonstrated for local-update and cluster simulations of simple lattice-spin
models \cite{block:10,weigel:10c}, for disordered systems
\cite{weigel:10a,lulli:15,navarro:16}, and for parallel tempering
\cite{weigel:10a,gross2011}. Here we introduce a GPU implementation of a
parallel-updating scheme~\cite{zierenberg2013} for multicanonical simulations and
discuss its performance for the case of the Ising model in two dimensions (2D). In
Sec.~\ref{secPmuca} we outline the multicanonical sampling method in its serial and
parallel versions, and discuss an efficient implementation of this approach on GPUs.
Section~\ref{secPerformance} is devoted to a detailed analysis of the performance of
the approach and its implementation for the reference problem of the 2D Ising
model. Finally, Sec.~\ref{secConclusion} contains our conclusions.

%%%%%%%%%%%%%%%%%%%%%%%%%%%%%%%%%%%%%%%%%%%%%%%%%%%%%%%%%%%%%%%%%%%%%%%%%%%%%%%%
%%%%%%%%%%%%%%%%%%%%%%%%%%%%%%%%%%%%%%%%%%%%%%%%%%%%%%%%%%%%%%%%%%%%%%%%%%%%%%%%
%%%%%%%%%%%%%%%%%%%%%%%%%%%%%%%%%%%%%%%%%%%%%%%%%%%%%%%%%%%%%%%%%%%%%%%%%%%%%%%%
\section{Parallel Multicanonical Algorithm\label{secPmuca}}

\subsection{Multicanonical sampling}

We confine ourselves here to a short introduction of the multicanonical
method~\cite{bergMuca, jankeMuca}, more details can be found in the literature. The
basic idea is to use additional weight factors to artificially enhance transition
states with low canonical probabilities, thus reducing free-energy barriers found in
the system. This feature makes the approach particularly suitable for the simulation
of first-order phase transitions and for systems with rugged free-energy
landscapes. In the standard approach, this results in a flat energy histogram and
thus an equal sampling of all energies, including minima as well as suppressed
transition states. The determination of appropriate weight factors requires an
iteration of runs. Since each iteration is performed for a fixed set of weights,
however, an unbiased reweighting to the canonical (or another) ensemble is possible
at all times.  This approach may be further generalized to other control parameters
such as external fields~\cite{jankeMuca, zierenberg2015bc}, bond and cluster numbers
\cite{wj:95a,weigel:10d}, overlap parameters~\cite{berg1998overlap}, or even
background obstacles~\cite{schoebl2011}.

For definiteness, consider the standard case of the inverse temperature
$\beta=1/k_BT$ as control parameter with the conjugate variable being the
conformational energy $E$. The canonical energy distribution is
\begin{equation}
P_{\beta}(E)=\frac{1}{Z_\beta}\Omega(E)e^{-\beta E},
\end{equation}
where $\Omega(E)$ is the density of states and
\begin{equation}
Z_{\beta}=\sum_{\{x_i\}}e^{-\beta E(\{x_i\})}=\sum_E \Omega(E)e^{-\beta E}
\end{equation}
is the canonical partition function. The sum can run equivalently over the
microscopic configurations $\{x_i\}$ or the energies $E$. In general, in a
multicanonical simulation we replace the Boltzmann factor $e^{-\beta E}$ with an
auxiliary weight function $W(E)$ such that
\begin{equation}
  Z_{\rm muca} 
  = \sum_{\{x_i\}} W\left(E\left(\{x_i\}\right) \right)
  =\sum_E \Omega(E) W(E).
\end{equation}
It is easy to see that energy states occur with the same probability, as requested
for a flat energy histogram, if the weight function becomes equal to the inverse of
the density of states $W(E) = \Omega^{-1}(E)$. Since this is in general not known in
advance, however, we need to iteratively adapt $W(E)$ to achieve the goal of a flat
histogram. It is straightforward to confirm that an estimator of the inverse density
of states (up to a multiplicative constant) from a simulation with weights
$W^{(n)}(E)$ and sampled (normalized) energy histogram $H^{(n)}(E)$ is
\begin{equation}
  \label{eq:weight_modification}
  W^{(n+1)}(E) \equiv \hat{\Omega}^{-1,(n)}(E) = W^{(n)}(E)/H^{(n)}(E).
\end{equation}
As the range of energies sampled increases with $n$, this approach will ultimately
converge to the desired weight function $W(E) = \Omega^{-1}(E)$. More sophisticated
schemes may combine neighboring histogram entries and the full statistics of previous
iterations~\cite{jankeMuca}, but we stick to the simple scheme here as we find it to
be more stable for the parallel application we are interested in. Once the weight
iteration has converged, which is typically checked by requiring that the
corresponding histogram satisfies a flatness criterion, the actual simulation in the
multicanonical ensemble (``production run'') is performed and all quantities of
interest are sampled. The final results can then easily be reweighted back to the
desired physical ensemble using the weight factors $W^{(\mathrm{final})}(E)$. In a
generalized formulation of the multicanonical method, the auxiliary weight becomes a
function of partial energy or order parameter. The following considerations are
completely independent of the updating scheme and equally applicable to generalized
formulations.

\subsection{Parallel multicanonical sampling}

\begin{figure}
  \includegraphics{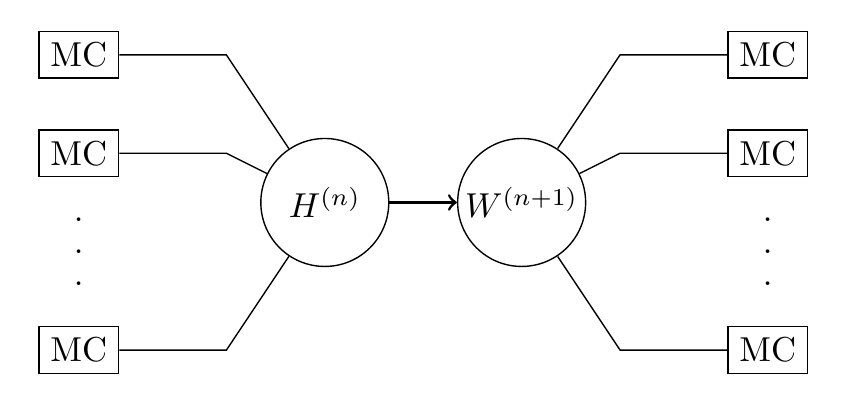}
  \caption{\label{figScheme}%
    Schematic representation of the parallel implementation of the multicanonical
    method. In each iteration $p$ independent Markov chains run in parallel with the
    same auxiliary weight function. The independent histograms are merged to a total
    histogram from which a new weight function is estimated and distributed onto the
    parallel walkers again. Adapted from Ref.~\cite{zierenberg2013}.}
\end{figure}

As discussed above, due to the dependence of each update on the global variable $E$,
domain decomposition is not a good option for generalized-ensemble
simulations. Parallel implementations therefore either use windows in the dependent
variable (here the energy) or independent or moderately dependent runs that combine
statistics~\cite{zierenberg2013,sugihara2009,slavin2010,ghazisaeidi2010}. Here, we
follow the scheme proposed in Ref.~\cite{zierenberg2013} and consider $p$ Markov
chain Monte Carlo simulations (``walkers'') with identical weight functions
$W^{(n)}_{i}=W^{(n)}$, $i=1,...,p$. Each simulation has the same, time-independent
stationary distribution determined by the weight $W^{(n)}$, and yields a histogram
$H^{(n)}_{i}(E)$. Since all histograms are (unnormalized) estimates of the same
probability distribution, we may add them directly to obtain the total histogram of a
given iteration \mbox{$H^{(n)}(E) = \sum_{i}H^{(n)}_{i}(E)$}.  $H^{(n)}(E)$ is then
used together with the current $W^{(n)}$ to determine the weight function $W^{(n+1)}$
for the next iteration from Eq.~\eqref{eq:weight_modification}, or it might be
alternatively determined by a more sophisticated weight iteration
scheme~\cite{jankeMuca}. The new weight function is again distributed to all $p$
walkers, and they perform new simulations with fixed weights
$W^{(n+1)}_{i} = W^{(n+1)}$. A schematic illustration of this parallel iteration is
shown in Fig.~\ref{figScheme}. Proceeding in this way the computational effort of the
iteration procedure may be efficiently distributed onto many cores, yielding the same
quality of results in a fraction of time. The production run can be parallelized
trivially in the same fashion, with a final accumulation of the sampled histograms.

As the histogram $H^{(n)}(E) = \sum_{i}H^{(n)}_{i}(E) $ is a result of independent
sampling processes, one might in fact argue that it is a better estimate than a
histogram from a single walker with the same total number of steps but affected by
autocorrelations. The situation is not quite as simple, however: while all walkers
have the same stationary distribution, still each simulation needs to perform
equilibration steps to become stationary. This is the case in each weight iteration
step as the starting configuration will be in equilibrium with respect to the
previous weight function only which is different from the current one. When
distributing the same number of updates over more and more walkers, a larger and
larger fraction of the runs needs to be spent on equilibration and can hence not be
used to estimate the histogram determining the weights for the next iteration. Note
that while in many scalar implementations of multicanonical simulations no
thermalization steps are used between iterations, for the large number of parallel
walkers employed here and the resulting small number of updates per iteration and
walker we find the equilibration steps to be crucial for achieving a stable parallel
procedure.  Recent applications of the parallel multicanonical method (on CPU
clusters) include studies of the Blume-Capel spin model in
2D~\cite{ZieFytWeiJanMal2017} and 3D~\cite{zierenberg2015bc}, lattice and off-lattice
particle
condensation~\cite{ZieWieJan2014,ZieJanPRE2015,NusZieBitJan2016,ZieSchJan2017},
continuum formulations of the aggregation process of
flexible~\cite{ZieSchJan2017,ZieMueSchMarJan2014} as well as
semiflexible~\cite{ZieJanEPL2015} polymers, the binding transition of grafted
flexible polymers~\cite{ZieThoJan2017}, the phase diagram of semiflexible
polymers~\cite{MarJan2016}, and the interplay of semiflexibility with the adsorption
propensity of polymers~\cite{AusZieJan2017}.  In all these different cases, the
method proved to be very robust and reliable to use routinely in day-by-day practical
work.

%%%%%%%%%%%%%%%%%%%%%%%%%%%%%%%%%%%%%%%%%%%%%%%%%%%%%%%%%%%%%%%%%%%%%%%%%%%%%%%%
%%%%%%%%%%%%%%%%%%%%%%%%%%%%%%%%%%%%%%%%%%%%%%%%%%%%%%%%%%%%%%%%%%%%%%%%%%%%%%%%

\subsection{Implementation on GPU\label{secPmucaGPU}}

In designing an implementation of the parallel multicanonical approach on GPUs, one
needs to take some basic architectural facts of such devices into account to achieve
decent performance. Of crucial importance here are the parallel computing model and
the hierarchy of memories. We restrict our attention and hence the discussion to
Nvidia GPUs, which we program using Nvidia CUDA, an extension of the C/C++
programming language for the GPU code. GPUs provide a hybrid environment, combining
elements of distributed and shared memory systems: threads in the same \emph{thread
  block} all have read and write access to a small but fast on-chip cache memory
which, for the GPUs considerd here, is at most $\SI{48}{KB}$.  Threads in a block all
reside on the same multiprocessor. Threads in different blocks have read and write
access to global memory only and need to use this channel for communication. For the
present problem, communication needs between threads are minimal and shared memory is
not explicitly required. We hence request shared memory to be used as L1 cache
instead where this option is available (using {\tt cudaDeviceSetCacheConfig()}).
Good performance heavily relies on the approach of {\em latency hiding\/}, a process
by which the GPU thread scheduler continuously puts threads that are waiting for data
accesses into a dormant state and activates other thread groups that are ready for
using the compute units. This works well as long as there is a large number of
available threads at any time, ideally exceeding the number of available compute
units by many times. Due to restrictions in the total number of resident threads and
thread blocks, the available registers, as well as the number of available
multiprocessors and cores per multiprocessor, a good division of the required threads
into thread blocks is not always easy to determine. To facilitate this task, the CUDA
toolkit provides an occupancy calculator spreadsheet, which we used to determine an
optimal block size of 256 threads for the devices used here. As is discussed below in
Sec.~\ref{secPerformance}, the optimal performance for the present problem is reached
if each multiprocessor is fully loaded with the maximally allowed number of resident
threads.

Probably the most important single optimization consideration relates to the locality
of memory accesses that can ensure {\em coalescence\/} of memory transfers: since
each access to global memory leads to the transfer of at least $128$ bytes (32 words,
a cache line), full utilization of the transfered data and hence the bus capacity is
only achieved if 32 consecutive threads (a {\em warp\/}) access memory from the same
128 bytes block simultaneously. This is achieved here via two tricks: as one thread
is assigned to each of the parallel walkers and hence works on its own copy of the
system (a spin model lattice in the examples discussed below), it is important to
choose the right memory arrangement of configurations. While it might seem natural to
place all of the spin configurations in a linear order next to each other, we put all
of the first spins next to each other, then all of the second spins etc. In this way
we ensure that if all walkers access the same lattice sites in their respective
replicas the corresponding memory accesses are fully coalesced.  If each walker
chooses lattice sites for spin flips independently at random, however, no coalescence
will typically be achieved and memory accesses remain highly non-local. This problem
can be circumvented by using the same random number sequence for selecting the spins
for all walkers, such that an update attempt simultaneously occurs at the same site
of each replica in a warp of walkers. The acceptance of updates using the Metropolis
criterion is then tested with a second stream of random numbers that is chosen
independently for each walker (as otherwise, of course, all of the walkers would be
highly correlated)\footnote{Although the synchronous choice of update sites formally
  introduces a correlation between walkers, the resulting simulation data are fully
  compatible statistically with results from simulations with independently chosen
  update sites which, however, run substantially less efficiently.}.

The use of high-quality random-number generators\linebreak{}(RNGs) is crucial in
order to prevent systematic biases from correlations, especially in the discrete
lattice models considered here (see, e.g., Ref.~\cite{ferrenberg:92}). Due to the
high frequency of random-number consumption and the relatively low arithmetic load of
the remaining calculations, the method of choice for the present problem is the
provision of an independent instance of RNG for each walker. The need to generate
large numbers of uncorrelated, parallel streams of random numbers as well as the
pressure on memory bandwidth exerted by RNGs with large state vectors lead to
somewhat different selection criteria for good RNGs for GPUs as compared to serial
applications on CPU~\cite{manssen:12,LevRNG}. For the present application we decided
for the stateless, counter-based RNG Philox proposed in Ref.~\cite{salmon:11}. This
has been thoroughly tested previously for lattice spin systems, and was shown to
combine high performance with excellent statistical properties~\cite{manssen:12}. The
counter-based nature means that no RNG state needs to be transferred over the bus
between global memory and compute units. The large number of independent sub-streams
can be labeled according to lattice site and walker, such that exactly the same
random numbers are used for corresponding decisions in all configurations
irrespective of the compute environment (GPU or CPU).\footnote{We checked a number of
  further generators and found that for the present application the performance of
  Philox is close to that of a simple, in-line 32-bit linear-congruential generator
  (with known poor statistical properties) which, by its simplicity and small state,
  can be regarded as a theoretical upper bound in RNG performance.} For details see
the discussions in Refs.~\cite{manssen:12,salmon:11}.

\begin{table*}[b]
  \caption{\label{tabHardware}%
    CPU and GPU hardware used for the comparisons with selected properties, including
    the clock speed, the number of total cores, the number of (next generation)
    streaming multiprocessors (SMX), the memory bandwidth and the power
    consumption (thermal design power TDP). Both GPUs are from the Kepler
    generation such that each SMX features 192 cores. 
  }
  \centering
  \begin{tabular}{l r r r r r r r}
    \hline
    & clock speed    & cores & SMX & memory bandwidth & peak perf. (SP) & TDP \\
    \hline
    CPU: $2\times$ Xeon E5-2640         & \SI{3072}{MHz} & 12     & -
                                                                                  &\SI{42.6}{GB/s}
                                                      & $2\times$ \SI{120}{GFlop/s}& $2\times$\SI{ 95}{W}\\
    GPU: Nvidia Tesla K20m (ECC)       & \SI{ 706}{MHz} & 2496  & 13  &\SI{
                                                                             208}{GB/s}
    & \SI{3.5}{TFlop/s} & \SI{225}{W}\\
    \phantom{GPU:} Nvidia GTX Titan Black & \SI{ 980}{MHz} & 2880  & 15  & \SI{
                                                                                336}{GB/s}
    & \SI{5.1}{TFlop/s} & \SI{250}{W}\\
    \hline
  \end{tabular}
\end{table*}

The actual Metropolis updates of spins according to the current weight function
$W(E)$ requires the evaluation of $\exp[\omega(E')-\omega(E)]$, where
$W(E') = \exp[\omega(E')]$ denotes the weight of the proposed configuration.  We are
working with logarithmic weights $\omega(E)=\ln W(E)$ to avoid numerical issues such
as overflows.  The weights are the same for all walkers, and it turns out to improve
performance to store them in a {\em texture\/} object, a GPU specific data structure
that is optimized for random read accesses. The arithmetically most time consuming
step of the update is the evaluation of the exponential function. We store the
weights in single precision and hence also use a single precision version of the
exponential function. GPUs feature special function units in hardware which provide
particularly high performance, however at the cost of a somewhat reduced
precision. To ensure identical results between the CPU and GPU versions of the code,
the example code provided does not use the hardware version \texttt{\_\_expf()},
however.

The weight iteration involves recording separate energy histograms for all walkers
that are then added up to derive the weight for the following iteration, cf.\ the
schematic sketch in Fig.~\ref{figScheme}. As the energies of walkers will typically
be very different from each other, memory accesses for incrementing histogram bins
will not be well coalesced. We considered three alternative implementations of the
energy histograms: (i) each walker has its own histogram, kept in global memory, and
these are added up in a separate kernel after each iteration to form the total
histogram; (ii) instead of storing histograms, each walker stores a list (time series)
of the energies encountered after each attempted spin flip, and the lists are
combined into a histogram in a separate kernel after each iteration; (iii) a single
histogram is kept in global memory and each walker increments energy bins using the
atomic operation \texttt{atomicAdd()}, which automatically resolves access conflicts,
such that a final addition of histograms becomes unnecessary. Variant (ii) avoids the
memory coalescence problems of the other implementations, but at the expense of
increased memory consumption and a more expensive histogram addition kernel. Overall,
however, we find by far the best performance for the atomic-update variant
(iii). There, in case of conflict a serialization of accesses must occur, thus slowing
down the code. Such events become rare in the limit of flat histograms as soon as the
number of possible energies becomes large against the number of cores on GPU.
This is only the case for systems larger than those studied here, but we still
find excellent performance of the atomic version as discussed below.

To assure a fair comparison of GPU and CPU performance we wrote the same optimized
code from scratch for both platforms reusing as much code as possible for both
implementations. As discussed above, this includes using exactly the same random
number streams for both versions. Performance comparison is a rather subtle task as
there are no clear-cut criteria according to which to select the hardware units to
use. We here chose to compare CPU nodes with two 6-core Xeon CPUs and hyper-threading
enabled (resulting in 24 threads) to nodes with one of two GPU devices, either the
high-end consumer card GTX Titan Black or the GPGPU card Tesla K20m, cf.\ the details
collected in Table~\ref{tabHardware}. Both cards are based on the Kepler architecture
and have rather similar (integer and single-precision) performance
characteristics. The Tesla K20m card provides significantly larger double-precision
performance as well as error-correcting code (ECC) to ensure memory integrity. As it
turns out, however, both of these features do not have any relevance for the program
discussed here. The CPU and GPU systems have similar power consumption and the CPU
nodes and the Tesla K20m node were roughly similarly expensive at the time of
purchase. A node with the consumer card Titan Black can be constructed at a
significantly cheaper price, however.

%%%%%%%%%%%%%%%%%%%%%%%%%%%%%%%%%%%%%%%%%%%%%%%%%%%%%%%%%%%%%%%%%%%%%%%%%%%%%%%%
%%%%%%%%%%%%%%%%%%%%%%%%%%%%%%%%%%%%%%%%%%%%%%%%%%%%%%%%%%%%%%%%%%%%%%%%%%%%%%%%
%%%%%%%%%%%%%%%%%%%%%%%%%%%%%%%%%%%%%%%%%%%%%%%%%%%%%%%%%%%%%%%%%%%%%%%%%%%%%%%%

\section{Performance\label{secPerformance}}

We test and benchmark our implementation by considering a standard reference problem,
the ferromagnetic, nearest-neighbor Ising model on a $L\times L$ patch of the square
lattice with Hamiltonian
\begin{equation}
  \mathcal{H} = -J\sum_{\langle i,j\rangle} s_i s_j,
\end{equation}
where $J=1$ is a coupling constant. Extensive analytical results are available for
this problem, turning it into an ideal test case~\cite{McCoyWu:book}. At
$\beta_0=\ln\left(1+\sqrt{2}\right)/2$, the system undergoes a second-order phase
transition. This is clearly not a prototypical problem for the multicanonical method,
since at criticality cluster updates surely outperform multicanonical
simulations. However, it is the standard test case with available analytic solutions
for the density of states~\cite{beale:96a}.

% Kullback-Leibler !!
For the convergence of the multicanonical iteration procedure we require a flat
histogram in energy space, i.e., in \mbox{$[-2L^2, 2L^2]$}. States are spaced at
intervals $\Delta E = 4$ and, excluding the forbidden states at $\pm(2L^2-4)$, there
is a total of $N_\textrm{bins} = L^2-1$ accessible energies.  To assess the flatness
of the histogram we consider the Kullback-Leibler divergence~\cite{kullback:51},
sometimes called relative entropy. It measures the similarity of two probability
distributions $P(x)$ and $Q(x)$ defined over the same domain and may be written for
discrete distributions as\footnote{Please note that common definitions of $d_k$ in
  the computer science literature consider the logarithm with base 2, i.e.,
  $\log_2$. In this case, $d_k^\prime=\sum_x P(x)\log_2 [P(x)/Q(x)]=d_k/\ln{2}$.  }
\begin{equation}
  \label{eq:kullback}
  d_k=\sum_x P(x)\ln \frac{P(x)}{Q(x)}.
\end{equation}
While $d_k \ge 0$ always, it is not symmetric under the exchange of distributions and
hence it is not a proper distance. Still, it serves very well as a convergence
criterion. Note that due to $\lim_{p_i\to 0} p_i \ln p_i = 0$ empty bins do not
contribute to $d_k$, which is a useful property for the present application.

In our implementation we set $P(E_i)=H(E_i)/N_\textrm{updates}$, where
$N_\textrm{updates}$ is the total number of measurement updates (i.e.,
attempted individual spin flips) in the current iteration, and $Q(E_i) = 1/w$, where $w=\max(w',10)$ and
$w'$ is the current width (in units of $\Delta E$) of the energy range covered in the
previous iteration.  During the initial iterations, the number of updates after the
equilibration phase is taken to be
\begin{equation}
  \label{eq:nupdates}
  N_\textrm{updates} = 6 w^z / W,
\end{equation}
where $W$ denotes the number of parallel walkers; the factor of 6 and the power of
$z=2.25$ are related to the autocorrelation and the average acceptance rate of the
multicanonical update and are justified in Ref.~\cite{zierenberg2013}.
Once $w = L^2+1$, i.e., the whole energy range is covered, we increase the number of
measurement updates by a factor of $1.1$ in each iteration.
The number of updates is always truncated to an integer value and increased by
$1$ (ensuring that $N_{\rm updates} \geq 1$.).
The simulation ends if after an iteration we find that $d_k < d_{k,{\rm flat}} = 10^{-4}$, which turned out
to ensure a very well converged weight function.

%thermalization
We found it crucial for arriving at stable results to also include a thermalization
phase (updates that are not recorded in the histogram) before each iteration with
$N_\textrm{therm}=f_{\rm therm} w$ attempted spin flips and $f_{\rm therm}=30$.  It is clear
that a weight iteration that samples states that are not drawn from the equilibrium
distribution for the weight function $W^{(n)}(E)$ will not yield a good estimate for
the weight function of the next iteration, leading to oscillatory behavior in the
iterations. This is discussed in more detail in \ref{appMeasures}, where we also
compare the behavior of the Kullback-Leibler divergence to that of the Chebyshev
distance. The latter is related to the more commonly used flatness criterion
restricting the maximal deviation from the mean.

In comparing the performance of CPU and GPU implementations of the parallel
multicanonical algorithm, we must take into account that the number of iterations is
in itself a random variable that, in particular, depends on the number $W$ of
parallel walkers. We hence first study the direct or ``hardware'' performance of
a single spin flip in the two codes to only then include the application
or ``software'' performance in a second step.

%%%%%%%%%%%%%%%%%%%%%%%%%%%%%%%%%%%%%%%%%%%%%%%%%%%%%%%%%%%%%%%%%%%%%%%%%%%%%%%%
%%%%%%%%%%%%%%%%%%%%%%%%%%%%%%%%%%%%%%%%%%%%%%%%%%%%%%%%%%%%%%%%%%%%%%%%%%%%%%%%
\subsection{Hardware performance\label{secHardware}}

\begin{figure}[t]
  \includegraphics{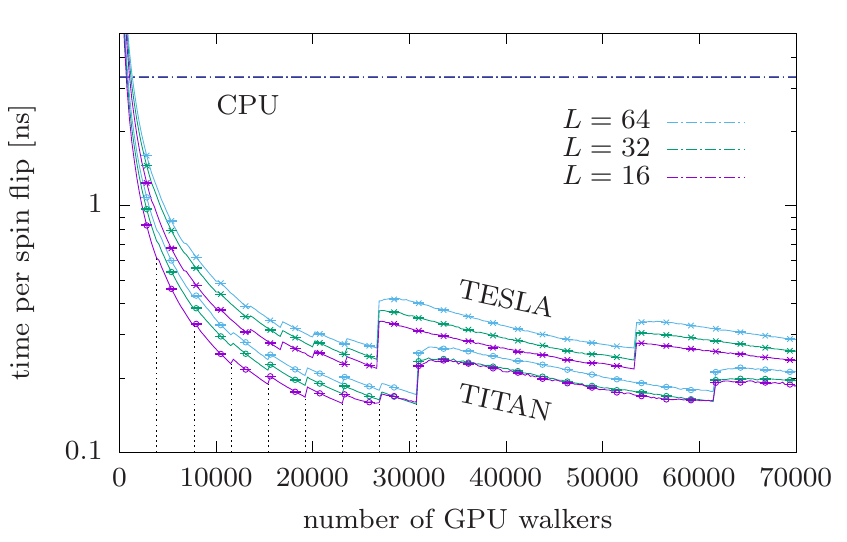}
  \caption{\label{figSpinflipTime}%
    Average time per spin flip in nanoseconds for the two GPUs as a function of the
    number of walkers, $W$, on GPU. Solid lines are for measurements in steps of
    $\Delta W = 256$, corresponding to one thread block; error bars are shown only
    for every tenth data point for clarity. The dashed vertical lines indicate
    multiples of one thread block per SMX for the Titan Black card, corresponding to
    $3840$ threads. The horizontal lines are reference times obtained from
    simulations on one CPU node with 12 cores, using 24 threads with hyper-threading.
  }
\end{figure}

To assess the performance of our codes on CPU and GPU for systems typically available
to users of current workstations and small to medium sized compute clusters, we chose
to compare one of two alternative Nvidia Kepler GPUs to a compute node with two Xeon
E5-2640 CPUs (for detailed specifications see Table~\ref{tabHardware}). The K20m is
from the Tesla range of professional GPU accelerator cards, featuring 13
next-generation streaming multiprocessors (SMX) with 192 cores each, totaling in 2496
CUDA cores. The ECC available on these cards on demand was turned on. The results are
contrasted to those of the high-end gaming card GTX Titan Black with 15 Kepler SMX
and a total of 2880 cores, featuring somewhat higher clock frequency, but no ECC and
lower double-precision performance as compared to the Tesla K20m. In practice, it
turns out that the lack of the latter two features is practically irrelevant for the
present application.

To single out the effect of hardware performance we measured the average spin-flip
time in nanoseconds for both GPUs and the CPU. To this end, we ran the simulations
for a range of different numbers of parallel walkers for a fixed number of
attempted spin flips, starting out from the known perfect multicanonical weights as derived from
the density of states~\cite{beale:96a}. Time measurements were only taken after the
equilibration phase.  The results for systems of linear sizes $L=16$, $L=32$, and
$L=64$ are shown in Fig.~\ref{figSpinflipTime}. Due to the parallelism being only
between walkers and not through domain decomposition of single spin lattices, the
results are only weakly dependent on system size, the effect being mostly related to
the changing degree of locality of memory accesses in incrementing the energy
histograms. As expected, we find somewhat smaller times for the Titan card with
higher clock frequency. Note, however, that this card performs dynamic frequency
scaling in a window below the maximum clock frequency to automatically control the
temperature and power consumption. To more clearly exhibit the algorithmic effects,
we added a few seconds of sleep after producing each data point, thus preventing the
card from clocking down.

The general trend of decreasing spin-flip times as the number of walkers is increased
is an effect of the latency hiding approach discussed above that leads to generally
improved performance as the cards are overloaded beyond the physical number of
cores. There is a hierarchy of two levels of steps in the data which are connected to
the commensurability of the loads to the available resources. One block of 256
threads per SMX corresponds to 3328 threads for the Tesla K20m with 13 SMX and 3840
threads for the Titan with 15 SMX. For any walker numbers in between these steps, the
SMX are unevenly loaded, leading to sub-optimal performance. The larger-scale steps
occur when the maximum number of eight resident 256-thread blocks per SMX is reached,
corresponding to $N_\mathrm{opt} = 8\times 3328 = 26\,624$ and
$N_\mathrm{opt} = 8\times 3840 = 30\,720$ threads for the Tesla K20m and Titan Black,
respectively. Beyond these optima, the surplus blocks need to run in a separate
round, again at sub-optimal uneven load of the SMX. We use these observations to
automatically set the number of walkers to be at the first and lowest minimum if the
user does not provide $W$ explicitly. This can be achieved using the kernel directive
\texttt{\_\_launch\_bounds\_\_}.

The straight line in Fig.~\ref{figSpinflipTime} is the reference time for our CPU
node, run with 24 threads on the dual Xeon nodes, which comes in at about
$\SI{3.3}{ns}$, practically independent of system size. The optimal spin-flip times
of $\SIrange{0.22}{0.27}{ns}$ for the Tesla K20m and $\SIrange{0.16}{0.17}{ns}$ for
the Titan Black result in effective hardware speedup factors up to $15$ and $21$,
respectively.  Note that in this performance comparison we use a whole CPU node with
12 cores as the reference. In the customary comparison to a single CPU core, these
numbers thus would have to be multiplied by a factor of 12, resulting in speedups of
180 and 252, respectively.  The factor of $\SIrange{1.4}{1.6}{}$ between the two
GPUs is compatible with the ratio of the specified peak performances of the cards as
indicated in Table~\ref{tabHardware} and is a result of the larger number of SMX
units and increased clock frequency of the Titan Black card as compared to the Tesla
K20m.

%%%%%%%%%%%%%%%%%%%%%%%%%%%%%%%%%%%%%%%%%%%%%%%%%%%%%%%%%%%%%%%%%%%%%%%%%%%%%%%%
%%%%%%%%%%%%%%%%%%%%%%%%%%%%%%%%%%%%%%%%%%%%%%%%%%%%%%%%%%%%%%%%%%%%%%%%%%%%%%%%
\subsection{Software performance\label{secSoftware}}

\begin{figure}[t]
  \includegraphics{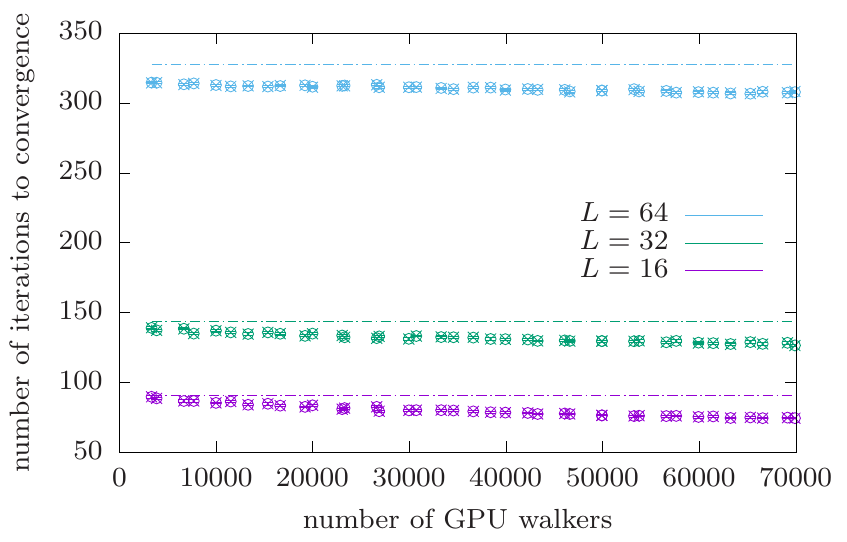}
  \caption{\label{figIter}%
    Average number of iterations until convergence for the parallel multicanonical
    method on GPU as a function of the number of walkers.  The dashed lines indicate
    the number of iterations for the CPU (24 walkers) for the three different system
    sizes.  Note that the number of iterations only depends on the number of walkers
    and the RNG seeds and hence is identical for runs on both GPUs.}
\end{figure}

We proceed with the algorithmic or software performance. To capture it, we
investigated the number of weight iterations required and the resulting total GPU
computing time until convergence as a function of the number of GPU walkers in steps
of the block size $256$. The number of iterations in this process is in itself a
random variable, and hence to derive meaningful results we averaged over $32$
independent runs performed with different RNG seeds. Error bars were obtained from
the fluctuations of the measurements. The parallelization employed here is via
independent copies of the system that all contribute to the sampling of the same
probability distribution. If in equilibrium, the independence of the copies should
improve the estimate due to a reduction in autocorrelations as compared to a purely
sequential process.

As outlined above, we employ an equilibration phase of $N_\mathrm{therm}\propto w$
updates (with width $w=L^2+1$ for the full energy range) at the beginning of each
iteration. This is a compromise in that it is not necessarily enough to sufficiently
equilibrate. As the number of walkers is increased, the number of histogram samples
$N_\mathrm{updates}$ is reduced according to Eq.~\eqref{eq:nupdates}, such that the
proportion of time spent on equilibration increases with $W$. As a result, it is seen
from our results summarized in Fig.~\ref{figIter} that the required number of
iterations mildly decreases with $W$ and seemingly settles. The behavior of the
required number of iterations depends significantly on the number $N_\mathrm{therm}$
of equilibration updates, and for shorter thermalization phases the required number
of iterations is found to increase with $W$ (not shown). As expected, the number of
iterations increases with the system size $L$ as this increases the range of possible
energy values to be covered by the walkers.

\begin{figure}[t]
  \includegraphics{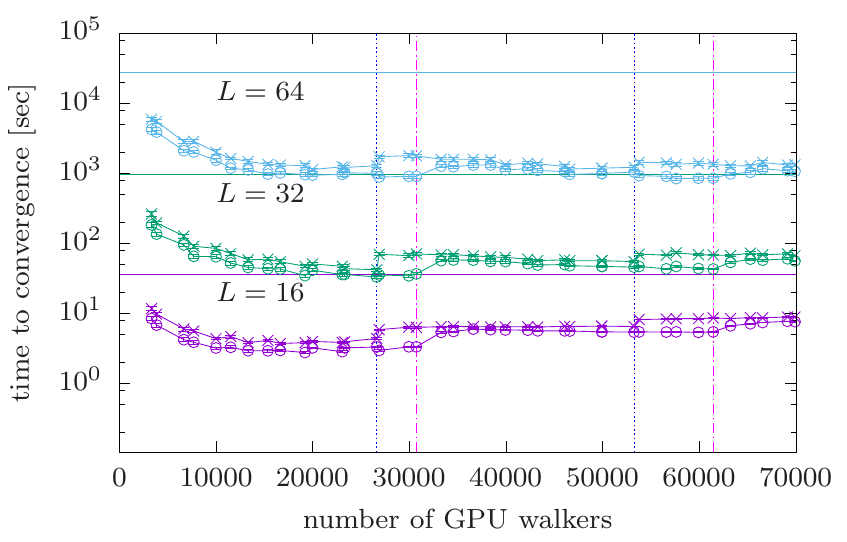}
  \caption{\label{figTime}%
    Average simulation time until convergence for the parallel multicanonical method
    on two different GPUs (crosses: Tesla K20m, circles: Titan Black) as a function
    of the number of walkers. The horizontal lines show the times obtained with the
    identical algorithm on one CPU node (24 threads using hyper-threading). Vertical
    dashed lines indicate the optimal occupancy as derived from
    Fig.~\ref{figSpinflipTime}.  }
\end{figure}

The total GPU computing time as a function of the number of GPU walkers is shown in
Fig.~\ref{figTime}. Horizontal lines indicate the average computation times on 12 CPU
cores (24 threads with hyper-threading). As expected, the total simulation time
clearly increases with system size. The GPU simulations finish in shorter time in the
full range of cases considered. Comparing the two different GPU environments
considered, we find that again the Titan Black card is consistently faster than the
Tesla K20m, as expected from the hardware specifications. The marked minima in the
total computational time indicated by vertical dashed lines coincide with the minima
in the hardware times shown in Fig.~\ref{figSpinflipTime} and are hence a result of
the optimal load of the cards achieved there. Although in some cases the higher-order
minima are not significantly worse than the first one, we recommend using the first
optimum as automatically chosen by our code, which can also be easily calculated
manually in the CUDA Occupancy Calculator spreadsheet as the product of ``total
number of multiprocessors'' and ``maximal threads per multiprocessor''.

\begin{figure}[t]
  \includegraphics{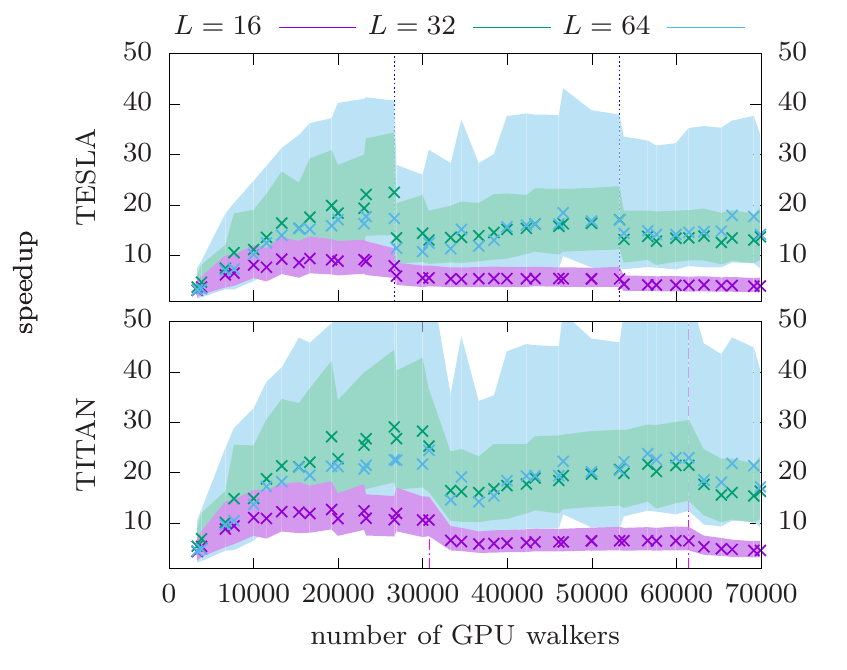}
  \caption{\label{figSpeedup}%
    Estimated speedup, i.e., reduction in wall-clock time required until convergence,
    for the parallel multicanonical method on different GPUs (Tesla K20m, Titan
    Black) as a function of the number of walkers. Vertical dashed lines
    indicate the optimal occupancies. The speedup is obtained compared
    to reference times from simulations on one CPU node (24 threads with
    hyper-threading). Data points mark the median speedup and the shaded areas
    indicate the confidence interval including $2/3$ of the data.  }
\end{figure}

Comparing the computational time to convergence for runs on GPU and CPU we can derive
a speedup factor, which is shown in Fig.~\ref{figSpeedup}. The average speedup is
achieved at the point of optimal occupancy and works out at around $20$ and $25$ for
the Tesla K20m and Titan Black cards, respectively. The relative size of these
speedups is again compatible with the relative peak performance of these
devices. These speedups are slightly larger than those found for the spin-flip times
(which were 15 and 21 for the K20m and Titan Black, respectively), which is an effect
of the slight decrease in the number of iterations with $W$ shown in
Fig.~\ref{figIter}. From an environmental perspective this means that if the CPU node
requires $2\times\SI{95}{W}=\SI{190}{W}$ for a given task, the Tesla K20m and Titan
Black GPUs would achieve compatible results for $\SI{11.3}{W}$ and $\SI{10.0}{W}$,
respectively, according to their thermal design power as indicated in
Table~\ref{tabHardware}. Note, however, that this does not take into account the
power consumption of the host CPU system in case of the GPUs as well as the power
consumption of components other than the CPU itself in the CPU node.

\begin{figure}[t]
  \includegraphics{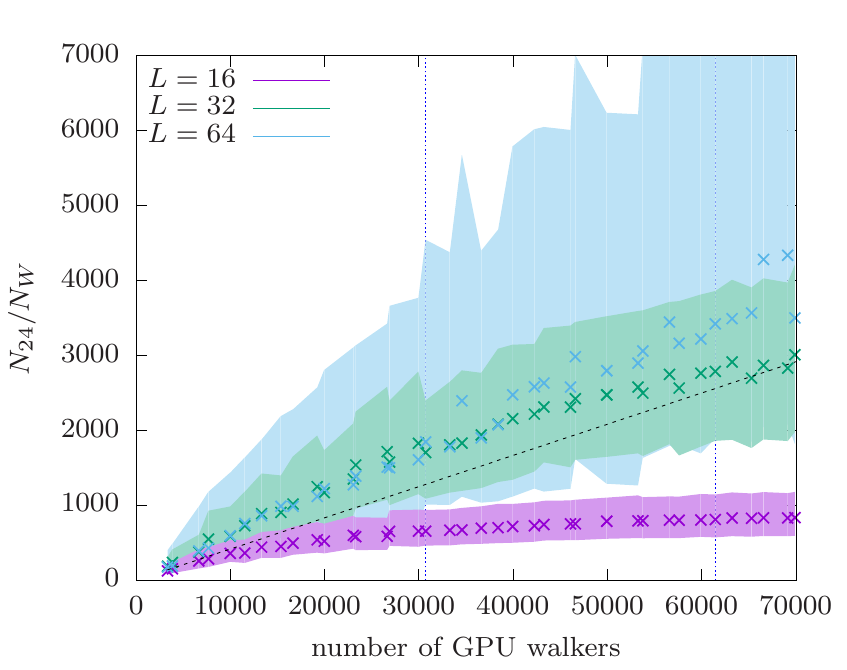}
  \caption{\label{figSpeedupStrong}%
    Estimated algorithmic speedup, i.e., ratio of the total number of updates per
    walker (including thermalization) until convergence for a simulation with $24$
    threads (corresponding to the CPU code) and a run with $W$ threads on the
    Titan Black GPU. Without
    statistical gain from independent walkers, the optimal ratio is
    $N_{24}/N_{W}=W/24$ indicated by the dashed black line. Data points
    mark the median speedup and the shaded areas indicate the confidence interval
    including $2/3$ of the data.%
    Vertical dashed lines indicate the optimal occupancies.
  }
\end{figure}

We can go one step further and consider a hardware-independent algorithmic speedup by
considering the ratio of the total number of updates (including thermalization)
performed per walker until convergence between the CPU and GPU implementations
\cite{zierenberg2013}. Figure~\ref{figSpeedupStrong} shows the corresponding strong
scaling plot for the Titan Black GPU. The optimal scaling, where each doubling of threads results in a
reduction of number of updates by a factor of two, is indicated by the dashed
line. We note that for intermediate values of $W$ we observe a super-linear speedup
that is an effect of the increased statistical independence of the parallel walkers
as compared to the serial run. For very large numbers of walkers, on the other hand,
we eventually expect an only sub-linear speedup due to the time spent in
thermalization. This effect is only clearly seen for $L=16$, however, while for
$L=32$ it just sets in at the edge of the range of $W$ considered
here, and it is not visible for $L=64$. For larger system sizes we expect the onset
of this effect to be shifted further towards larger numbers of parallel walkers. This
clearly indicates that for usual system sizes the method is efficiently applicable
for computations on massively parallel machines with many tens of thousands of cores.

%%%%%%%%%%%%%%%%%%%%%%%%%%%%%%%%%%%%%%%%%%%%%%%%%%%%%%%%%%%%%%%%%%%%%%%%%%%%%%%%
%%%%%%%%%%%%%%%%%%%%%%%%%%%%%%%%%%%%%%%%%%%%%%%%%%%%%%%%%%%%%%%%%%%%%%%%%%%%%%%%
\subsection{Verification of physical results}
\newcommand{\estLnOmega}{\overline{\ln\Omega}}

\begin{figure}[t]
  \includegraphics{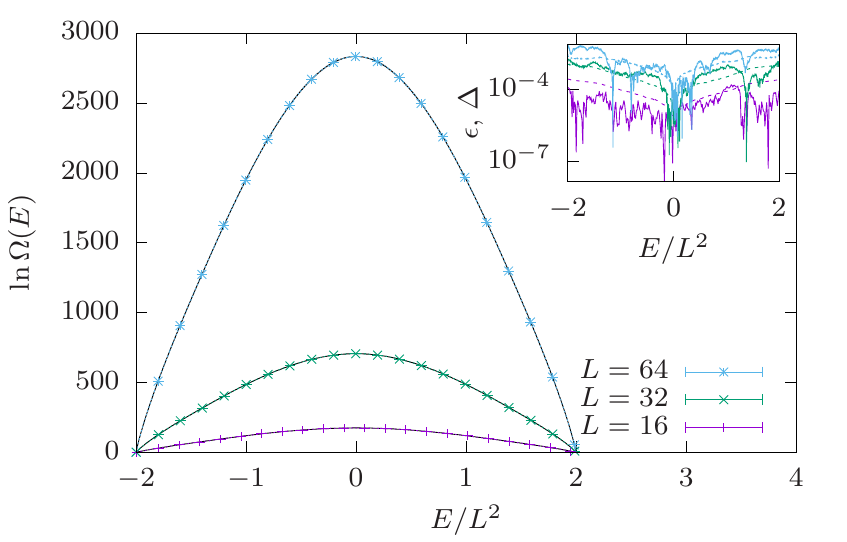}
  \caption{\label{figDOS}%
    Estimate of the logarithm of the density of states of the 2D Ising model from
    massively parallel multicanonical simulations on the Titan GPU card with
    $30\,720$ walkers and a fixed runtime of one hour a final production run
    performed for several system sizes. The solid lines show the exact solution
    according to Ref.~\cite{beale:96a}, which agrees well with the estimates within
    error bars. The inset shows the jackknife errors,
    $\epsilon=\epsilon(\estLnOmega(E))$, as dashed lines and absolute deviations from
    the exact solution, $\Delta=|\estLnOmega(E)-\ln\Omega(E)|$, as solid lines.%
  }
\end{figure}

To demonstrate the correctness and proper convergence of the method, we compare the
final estimate of the density of states resulting from a production run with fixed
weights to the known exact result~\cite{beale:96a}.  The estimate of the density of
states is obtained from a final production run with a total runtime of
$\SI{1}{h}$. Here, we restricted ourselves to the Titan GPU card with $30\,720$ GPU
walkers, which is the theoretically optimal choice as discussed above.  We use the
function of the code that is employed for the weight iterations also in the
production run, with the only difference that the resulting histogram is not used to
define new weights. To allow for the estimation of statistical errors, we divide the
total number of attempted spin flips into $100$ individual calls to the multicanonical simulation
function and use the resulting individual histograms to determine error bars from a
jackknife analysis~\cite{efron:82}.  Thus, a single simulation with parallel walkers
allows us to determine an estimate of the density of states and an estimate of the
error for each energy bin. The result is shown in Fig.~\ref{figDOS}. As seen in the
main panel, there is excellent agreement of the estimate $\estLnOmega(E)$, shown in
dashed colored lines with occasional error bars, to the exact result $\ln\Omega(E)$,
shown as black lines. In order to fix the overall normalization, we ensure that the
total number of states represented by $\exp[\estLnOmega(E)]$ is $2^{L^2}$. All
calculations are directly performed for the logarithm of $\Omega$ to avoid numerical
overflows. As a result, the error bars shown are errors of the logarithm of the
density of states.

A more detailed quantitative comparison is provided by the absolute deviation from
the exact result, $\Delta=|\estLnOmega(E)-\ln\Omega(E)|$, in comparison to the
jackknife error, $\epsilon=\epsilon(\estLnOmega(E))$, as shown in the inset of
Fig.~\ref{figDOS}. We observe the best agreement, i.e., the smallest 
deviation, around the center of the energy range. Here also the statistical error is
smallest. This is a natural consequence of the vastly larger number of available
states around $E=0$ as compared to the case of very large or very small energies. The
ground state, for instance is only twofold degenerate. These observations
notwithstanding, the statistical accuracy observed is very good and it can be
systematically improved by using even more statistics in the production run.
In general, we recommend to use the multicanonical weights to sample the quantities
of interest in a production run. This guarantees a known exponential convergence
behavior to equilibrium and allows for the reliable estimation of statistical
errors. This is in contrast to the approach of using the estimated weights and
density of states directly to estimate quantities such as the average energy and
specific heat, where the analysis of systematic and statistical errors is much more
difficult.
We repeated the production run also for the Tesla K20m with $26\,624$ walkers and a
full CPU compute node with $24$ walkers. We define the average statistical gain as
the ratio between the squared statistical errors estimated for the density of states,
$\epsilon^2_{\rm cpu}/\epsilon^2_{\rm gpu}$. We find average statistical gains of
about $17$ and $23$ for the Tesla K20m and the Titan Black cards, respectively, which
are consistent with the speedups found before for the weight iteration.

%%%%%%%%%%%%%%%%%%%%%%%%%%%%%%%%%%%%%%%%%%%%%%%%%%%%%%%%%%%%%%%%%%%%%%%%%%%%%%%%

%%%%%%%%%%%%%%%%%%%%%%%%%%%%%%%%%%%%%%%%%%%%%%%%%%%%%%%%%%%%%%%%%%%%%%%%%%%%%%%%
\section{Conclusions}
\label{secConclusion}

We demonstrated the suitability of parallel multicanonical simulations for massively
parallel architectures. As a test platform we considered graphics cards, which play
an increasingly important role in high-performance computing at all scales, in
particular due to their favorable relation of performance to price and power
consumption. To be representative of typical installations accessible to users, we
used two Nvidia GPUs, one from the consumer series (GTX Titan Black) and one
professional computing card (Tesla K20m).  In comparison to a full CPU node with 12
cores (and using 24 hyper-threads) we find a significant speedup if the card is
sufficiently occupied, with a median of around 25 for the GTX Titan Black and 20 for
the Tesla K20m GPU. For the perhaps more standard comparison to a single CPU core
these speedups need to be multiplied by a factor of 12, resulting in improvement
factors of 300 and 240, respectively.  Such speedups exceed the improvement expected
from the measured hardware performance, which is the result of an additional
algorithmic advantage due to the statistical independence of the walkers. This effect
is also reflected in a reduction of the number of required updates until
convergence. The physical result, namely an equilibrium estimate of the density of
states, is found to be in statistical agreement with the exact solution up to high
precision.

An efficient parallelization of Markov chain approaches is ultimately limited by the
need to equilibrate each parallel walker, since the thermalization itself cannot be
parallelized. For the present implementation of parallel multicanonical simulations
applied to at least 1000 spins, however, we find that this effect sets in only beyond
the maximum number of 70\,000 threads studied here. For realistic system sizes we
hence expect good scaling up to at least $10^5$ cores, demonstrating that such
schemes are applicable for calculations on the largest computational resources
available, including non-GPU hardware such as BlueGene/Q, Cray, or other massively
parallel compute cluster architectures.

\section*{Acknowledgments}
We would like to thank Marco Mueller for fruitful discussions. Part of this work has
been financially supported by the Deutsche Forschungsgemeinschaft (DFG) through
SFB/TRR 102 (project B04) and under Grant No.\break JA~483/31-1, the Leipzig Graduate
School of Natural Sciences ``BuildMoNa'', the Deutsch-Franz\"osische Hochschule
(DFH-UFA) through the Doctoral College ``${\mathbb L}^4$'' under \allowbreak Grant
No.~CDFA-02-07, and the EU through the IRSES network DIONICOS under contract
No.~PIRSES-GA-2013-612707.
The authors gratefully acknowledge a Research Grant of the Royal Society for the
simulation of disordered systems.

\appendix
\section{Measures of Convergence}
\label{appMeasures}
\setcounter{figure}{0} 

\begin{figure}[t]
  \includegraphics{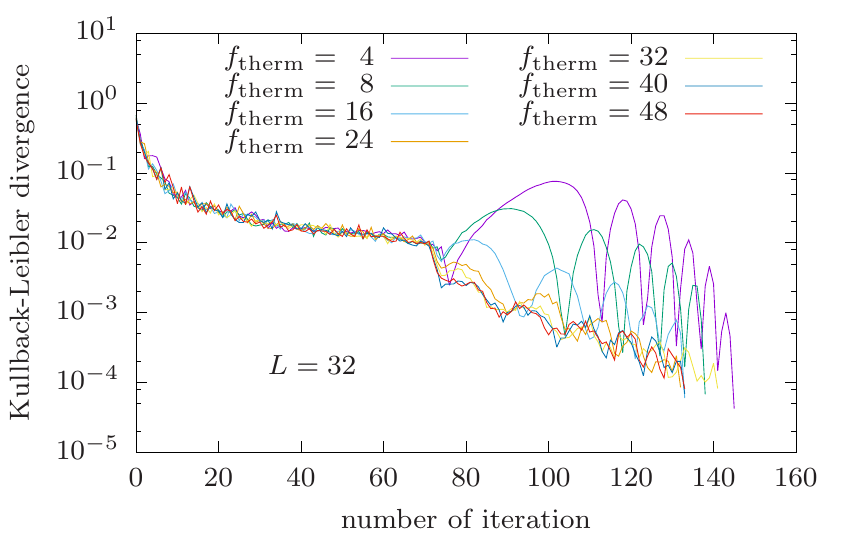}
  \includegraphics{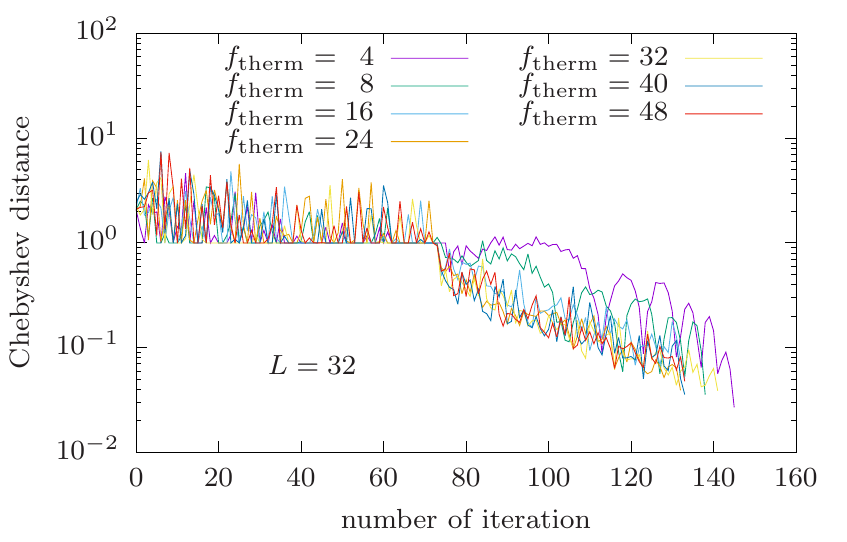}
  \caption{\label{figConvergence}%
    Effect of the number of thermalization sweeps $f_{\rm therm}$ on the convergence
    of the multicanonical iteration (Tesla K20m, $W=26\,624$) measured with the
    Kullback-Leibler divergence (top) or the Chebyshev distance (bottom) for an
    $L=32$ system.  The strong oscillations for small numbers of thermalization
    updates may be attributed to non-equilibrium artifacts which are carried along in
    the weight estimation.  }
\end{figure}

In Fig.~\ref{figConvergence} we present exemplary results from the multicanonical
weight iteration on the Tesla K20m card with $W=26\,624$ walkers showing two
different quantities used to assess convergence. The protocol used is described above
in Sec.~\ref{secPerformance}, using an initial thermalization phase of
$N_{\rm therm}=f_{\rm therm} w$ attempted spin flips followed by $N_\mathrm{updates}$ updates to the
histograms. According to \eqref{eq:nupdates} $N_\mathrm{updates}$ is systematically
increased once $w=L^2+1$ by multiplying it by a factor $1.1$ after each
iteration. The Kullback-Leibler divergence is defined in Eq.~\eqref{eq:kullback} and
once this measure falls below $10^{-4}$ the simulation aborts. In addition, we show
the Chebyshev distance $d_{c}$ defined as the maximal deviation of the histogram
$H(x)$ from the expected value of a flat distribution,
$m=N_{\rm updates}/N_{\rm bins}$
\begin{equation}
  d_c = \max_x(|H(x)/m-1|).
\end{equation}
This is closely related to common measures of a maximal deviation from the mean in
the context of generalized-ensemble methods. The initial lower bound $d_c\geq1$
observed from the lower panel of Fig.~\ref{figConvergence} is a result of almost
empty histogram bins with $H(x)\approx0$. From the data shown it is apparent that a
convergence criterion $d_c < d_{c,\mathrm{flat}}=0.05$ for the Chebyshev distance
would lead to a stopping time comparable to that resulting from the Kullback-Leibler
criterion.

Both measures are quite sensitive to the behavior at the boundaries of the energy
range discovered at any given time and the effect of empty bins. Note, however, that
the Kullback-Leibler divergence automatically assigns zero weight to such bins. After
the full range of energies has been discovered (here around 70 iterations), the number of
updates $N_{\rm updates}$ is increased by a factor of $1.1$ in each iteration, leading to a pronounced
change of behavior of the development of the convergence metrics at this point. We
notice rather strong oscillations in $d_k$ and $d_c$ for short thermalization times,
which we attribute to the presence of non-equilibrium artefacts in the recorded
histograms, leading to distorted weight functions for the following iterations. These
are carried over to successive iterations by the weight update and for low statistics
per walker they may take a long time to be corrected. In fact, one can easily imagine
that a sufficiently strong correction in the weights would require a sizable subset
of walkers to transit out of (or into) now less favorable (more favorable)
regimes. If the short thermalization pass does not allow for this to happen, the next
iteration records again too many (or too few) signals in these regimes, leading to
even stronger weight distortion.  This explains the oscillatory behavior.

We note that from general random-walk arguments, one might deduce a scaling of
$N_{\rm therm}\propto w^z$, but we found such a prescription to devote too much time
to thermalization. Hence, we chose the heuristic approach with
$N_{\rm therm}=f_{\rm therm} w$ and tuned $f_{\rm therm}$ to minimize
oscillations. It is clear that this tuning leads to model specific parameters, and
for optimal performance it would need to be repeated for different models to be
considered.

%\section*{\refname}


\begin{thebibliography}{10}
\bibitem{moore:65}
  G.E.~Moore, Electronics 38 (1965) 114.
\bibitem{mccool:12}
  M.~McCool, J.~Reinders, A.~Robison, Structured Parallel Programming: Patterns for Efficient Computation, Morgan
  Kaufman, Waltham, MA, 2012.
\bibitem{weigel:10b}
  M.~Weigel, Phys.~Rev.~E 84 (2011) 036709.
% \bibitem{swendsen:86}
%   R.H. Swendsen, J.S.~Wang, Phys.~Rev.~Lett. 57 (1986) 2607.
% \bibitem{geyer:91}
%   \bibinfo{author}{C.~J. Geyer},
%   in: Computing Science and Statistics: Proceedings of the 23rd Symposium on the Interface, American
%   Statistical Association, New York, 1991, p. 156.
\bibitem{hukushima:96a}
  K.~Hukushima, K.~Nemoto, J.~Phys.~Soc.~Jpn. 65 (1996) 1604.
\bibitem{bergMuca}
  B.A.~Berg, T.~Neuhaus, Phys.~Lett.~B 267 (1991) 249;
  Phys.~Rev.~Lett. 68 (1992) 9.
\bibitem{jankeMuca}
  W.~Janke, Int.~J.~Mod.~Phys.~C 03 (1992) 1137;
  Physica A 254 (1998) 164;
  Histograms and All That,
  in: Computer Simulations of Surfaces and Interfaces, NATO Science Series,
  Vol. 114, edited by B.~D\"unweg, D.P.~Landau, and A.I.~Milchev, Kluwer,
  Dordrecht, 2003, pp. 137--157;
  and in: Computational Many-Particle Physics,
  edited by H.~Fehske, R.~Schneider, and A.~Weiße,
  Lect.~Notes~Phys. 739, Springer, Berlin, 2008, pp. 79-140.
\bibitem{wang2001}
  F.~Wang, D.P.~Landau, Phys.~Rev.~Lett. 86 (2001) 2050;
  Phys.~Rev.~E 64 (2001) 056101.
\bibitem{schulz:03}
  B.J.~Schulz, K.~Binder, M.~M{\"{u}}ller, D.P.~Landau, Phys.~Rev.~E 67 (2003) 067102.
\bibitem{weigel:11}
  M.~Weigel, T.~Yavors'kii, Physics Procedia 15 (2011) 92.
\bibitem{zierenberg2013}
  J.~Zierenberg, M.~Marenz, W.~Janke, Comput.~Phys.~Commun. 184 (2013) 1155;
  Physics Procedia 53 (2014) 55.
\bibitem{vogel:13}
  T.~Vogel, Y.W.~Li, T.~W{\"{u}}st, D.P. Landau, Phys.~Rev.~Lett. 110 (2013) 210603.
\bibitem{owens:08}
  J.D.~Owens, M.~Houston, D.~Luebke, S.~Green, J.E.~Stone, J.C.~Phillips,
  in: Proceedings of the IEEE, Institute of Electrical and Electronics Engineers,
  New York, 2008, p. 879.
\bibitem{weigel2012}
  M.~Weigel, A.~Arnold, P.~Virnau (eds.), Computer Simulations
  on Graphics Processing Units, Eur.~Phys.~J. -- Special Topics 210,
  Springer, Heidelberg, 2012.
\bibitem{block:10}
  B.~Block, P.~Virnau, T.~Preis, Comp.~Phys.~Commun. 181 (2010) 1549.
\bibitem{weigel:10c}
  M.~Weigel, Comput.~Phys.~Commun. 182 (2011) 1833.
\bibitem{weigel:10a}
  M.~Weigel, J.~Comp.~Phys. 231 (2012) 3064.
\bibitem{lulli:15}
  M.~Lulli, M.~Bernaschi, G.~Parisi, Comput.~Phys.~Commun. 196 (2015) 290.
\bibitem{navarro:16}
  C.A.~Navarro, W.~Huang, Y.~Deng, Comput.~Phys.~Commun. 205 (2016) 48.
\bibitem{gross2011}
  J.~Gross, W.~Janke, M.~Bachmann, Comput.~Phys.~Commun. 182 (2011) 1638;
  Physics Procedia 15 (2011) 29.
\bibitem{zierenberg2015bc}
  J.~Zierenberg, N.G.~Fytas, W.~Janke, Phys.~Rev.~E 91 (2015) 032126.
\bibitem{wj:95a}
  W.~Janke, S.~Kappler, Phys.~Rev.~Lett. 74 (1995) 212.
\bibitem{weigel:10d}
  M.~Weigel, Physics Procedia 3 (2010) 1499.
\bibitem{berg1998overlap}
  B.A.~Berg, W.~Janke, Phys.~Rev.~Lett. 80 (1998) 4771.
\bibitem{schoebl2011}
  S.~Sch{\"o}bl, J.~Zierenberg, W.~Janke, Phys.~Rev.~E 84 (2011) 051805.
\bibitem{sugihara2009}
  T.~Sugihara, J.~Higo, H.~Nakamura, J.~Phys.~Soc.~Jpn. 78 (2009) 074003.
\bibitem{slavin2010}
  V.V.~Slavin, Low~Temp.~Phys. 36 (2010) 243.
\bibitem{ghazisaeidi2010}
  A.~Ghazisaeidi, F.~Vacondio, L.A.~Rusch, J.~Lightwave~Technol. 28 (2010) 79.
% additions from W. Janke
\bibitem{ZieFytWeiJanMal2017}
  J.~Zierenberg, N.G.~Fytas, M.~Weigel, W.~Janke, A.~Malakis, Eur.~Phys.~J. -- Special Topics 226 (2017) 789.
%\bibitem{ZieFytJan2015}
%  J.~Zierenberg, N.G.~Fytas, W.~Janke, Phys.~Rev.~E 91, (2015) 032126.
\bibitem{ZieWieJan2014}
  J.~Zierenberg, M.~Wiedenmann, W.~Janke, J.~Phys: Conf.~Ser. 510 (2014) 012017.
\bibitem{ZieJanPRE2015}
  J.~Zierenberg, W.~Janke, Phys.~Rev.~E 92 (2015) 012134.
\bibitem{NusZieBitJan2016}
  A.~Nußbaumer, J.~Zierenberg, E.~Bittner, W.~Janke, J.~Phys.: Conf.~Ser. 759, (2016) 012009.
\bibitem{ZieSchJan2017}
  J.~Zierenberg, P.~Schierz, W.~Janke, Nat.~Commun. 8 (2017) 14546.
\bibitem{ZieMueSchMarJan2014}
J.~Zierenberg, M.~Mueller, P.~Schierz, M.~Marenz, W.~Janke, J.~Chem.~Phys. 141 (2014) 114908.
\bibitem{ZieJanEPL2015}
  J.~Zierenberg, W.~Janke, Europhys.~Lett. 109 (2015) 28002.
\bibitem{ZieThoJan2017}
  J.~Zierenberg, K.~Tholen, W.~Janke, Eur.~Phys.~J. -- Special Topics 226 (2017) 683.
\bibitem{MarJan2016}
  M.~Marenz, W.~Janke, Phys.~Rev.~Lett. 116 (2016) 128301.
\bibitem{AusZieJan2017}
  K.~Austin, J.~Zierenberg, W.~Janke, Macromolecules 50 (2017) 4054.
% end additions from W. Janke
\bibitem{ferrenberg:92}
  A.M.~Ferrenberg, D.P.~Landau, Y.J.~Wong, Phys.~Rev.~Lett. 69 (1992) 3382.
\bibitem{manssen:12}
  M.~Manssen, M.~Weigel, A.K.~Hartmann, Eur.~Phys.~J.~Special~Topics 210 (2012) 53.
\bibitem{LevRNG}
  L.Yu.~Barash, L.N.~Shchur, Comput.~Phys.~Commun. 185 (2014) 1343.
\bibitem{salmon:11}
  J.K.~Salmon, M.A.~Moraes, R.O.~Dror, D.E.~Shaw,
  in: Proceedings of 2011 International Conference for High Performance Computing, Networking, Storage and Analysis,
  ACM, New York, 2011.
\bibitem{McCoyWu:book}
  B.M.~McCoy, T.T.~Wu, The Two-Dimensional Ising Model, Harvard University Press,
  Cambridge, 1973.
\bibitem{beale:96a}
  P.D.~Beale, Phys.~Rev.~Lett. 76 (1996) 78.
\bibitem{kullback:51}
  S.~Kullback, R.A.~Leibler, Ann.~Math.~Stat. 22 (1951) 79.
\bibitem{janke:92}
  W.~Janke, B.A.~Berg, M.~Katoot, Nucl.~Phys.~B 382 (1992) 649.
\bibitem{efron:82}
  B.~Efron, The Jackknife, the Bootstrap and Other Resampling Plans, Society for Industrial and
  Applied Mathematics [SIAM], Philadelphia, 1982.
\end{thebibliography}
\end{document}